\documentclass[a4paper,11pt]{article}
\usepackage{pos}
\usepackage{subcaption}

\usepackage{graphicx}
\usepackage{dsfont}
\usepackage{dcolumn}
\usepackage{multirow}
\usepackage{array}
\usepackage{url}
\usepackage{orcidlink}

\newcommand{\beq}{\begin{equation}}
\newcommand{\eeq}{\end{equation}}
\newcommand{\beqs}{\begin{eqnarray}}
\newcommand{\eeqs}{\end{eqnarray}}

\newcommand{\Tr}{{\rm Tr}}

\newcommand{\dd}{\mbox{d}}

\newcommand{\be}{\begin{equation}}
\newcommand{\ee}{\end{equation}}
\newcommand{\ba}{\begin{array}}
\newcommand{\ea}{\end{array}}

\newcommand{\orcidauthorPIAI}{0000-0002-2251-0111} 
\newcommand{\orcidauthorFATEMIABHARI}{0000-0003-1369-6505} 
\newcommand{\orcidauthorELANDER}{0000-0001-6348-8021}

\title{\boldmath Progress on Holographic Vacuum Misalignment}

\author*[a]{Ali Fatemiabhari\,\orcidlink{\orcidauthorFATEMIABHARI}}
\author{Daniel Elander,\,\orcidlink{\orcidauthorELANDER}}
\author[a]{Maurizio Piai\,\orcidlink{\orcidauthorPIAI}}

\affiliation[a]{Department of Physics, Faculty of Science and Engineering, 
Swansea University, Singleton Park, Swansea, United Kingdom}

\emailAdd{2127756@swansea.ac.uk}
\emailAdd{m.piai@swansea.ac.uk}
\emailAdd{daniel.elander@gmail.com}

\abstract{We summarise  highlights from an  ongoing research programme that aims, in the long run, at 
the ambitious goal of building a realistic, complete holographic composite-Higgs model.
This contribution focuses on vacuum misalignment, by showing how to unify
its description, as a phenomenon 
 arising from weak coupling considerations,
in the holographic description of a strongly coupled field theory in terms of a dual gravity theory.
This is achieved by a non-trivial treatment of boundary-localised terms in the gravity action.
The gravity backgrounds considered are completely regular and smooth.
We provide numerical examples showing that the  mass spectrum of particles in the four-dimensional theory is free of pathologies, and 
that a small hierarchy arises naturally, between the light states that, in this simplified set up, are analogous to  
the standard-model particles, and all the other, new composite states emerging in the strongly coupled theory.
}

\FullConference{The 41th International Symposium on Lattice Field Theory (Lattice 2024)\\
July 28th - August 3rd, 2024\\
University of Liverpool\\}

\begin{document}
\maketitle

\section{Introduction}

Composite Higgs Models (CHMs), in which the fields in the Higgs doublet of the Standard Model (SM) emerge as a set of Pseudo-Nambu-Goldstone-Bosons (PNGBs)~\cite{Kaplan:1983fs,Georgi:1984af,Dugan:1984hq}, in  a more fundamental, strongly coupled, confining  field theory,  offer a promising framework within which to address some of the big open questions of modern particle physics~\cite{Panico:2015jxa,Witzel:2019jbe,Cacciapaglia:2020kgq} (see also Refs.~\cite{Ferretti:2013kya,Ferretti:2016upr,Cacciapaglia:2019bqz}), such as the electroweak (big or small) hierarchy problem. The  new physics sector is endowed with an approximate global symmetry, described by a Lie group, $G$, broken to a subgroup, $H$, both by explicit symmetry breaking terms, as well as the emergence of composite condensates. At low energies the theory can be replaced by an Effective Field Theory (EFT), in which the PNGBs are described by fields taking values in the coset, $G/H$, along the same lines as the chiral Lagrangian. The distinctive feature of CHMs is that while the electroweak gauge group is embedded as a subgroup of $G$,  the presence of a perturbative instability, itself originating from the coupling to SM fields, induces misalignment with the vacuum~\cite{Peskin:1980gc}, and electroweak symmetry breaking.

As the phenomenology, in particular
 the mass spectrum of the composite states,
 is determined by the underlying 
 strongly coupled
 dynamics,
  it is natural to study it on the lattice.\footnote{Pertinent numerical lattice calculations exist
in theories with gauge group $SU(2)$~\cite{
Hietanen:2014xca,Detmold:2014kba,Arthur:2016dir,Arthur:2016ozw,
Pica:2016zst,Lee:2017uvl,Drach:2017btk,Drach:2020wux,Drach:2021uhl,Bowes:2023ihh},
 $Sp(4)$~\cite{Bennett:2017kga,
Bennett:2019jzz,Bennett:2019cxd,Bennett:2020hqd,Bennett:2020qtj,
Bennett:2022yfa,Bennett:2022gdz,Bennett:2022ftz,
Bennett:2023wjw,Bennett:2023gbe,Bennett:2023mhh,
Bennett:2023qwx,Bennett:2024cqv,Bennett:2024wda,
Maas:2021gbf,Zierler:2021cfa,
Kulkarni:2022bvh,Zierler:2022qfq,Zierler:2022uez,Bennett:2023rsl,Dengler:2023szi,Dengler:2024maq}
and $SU(4)$~\cite{Ayyar:2017qdf,Ayyar:2018zuk,Ayyar:2018ppa,
 Ayyar:2018glg,Cossu:2019hse,Lupo:2021nzv,
 DelDebbio:2022qgu,Hasenfratz:2023sqa}.
Results for the  $SU(3)$ theory with $N_f=8$ Dirac fermions~\cite{
LatKMI:2014xoh,
Appelquist:2016viq,
LatKMI:2016xxi,
Gasbarro:2017fmi,
LatticeStrongDynamics:2018hun,LSD:2023uzj,LatticeStrongDynamics:2023bqp,Ingoldby:2023mtf}
have  been reinterpreted in terms of new CHMs, embedded in the dilaton EFT framework~\cite{Appelquist:2020bqj,
Appelquist:2022qgl}---see also Refs.~\cite{Vecchi:2015fma,Ma:2015gra,BuarqueFranzosi:2018eaj}}
Unfortunately, for the minimal CHM, based upon the $SO(5)/SO(4)$ coset,
the low-energy spectrum of which consists only of the known SM fields, 
 a simple formulation, amenable to numerical lattice
 studies is not known (see Ref.~\cite{Caracciolo:2012je}). Moreover,  the aforementioned vacuum misalignment phenomenon is perturbative in nature, and its lattice treatment unwieldy.
Gauge-gravity dualities~\cite{
Maldacena:1997re,Gubser:1998bc,Witten:1998qj,
Aharony:1999ti} offer a promising alternative way to address calculability. Well known realisations of CHMs based on the  $SO(5)/SO(4)$ coset~\cite{Contino:2003ve,
Agashe:2004rs,
Agashe:2005dk,
Agashe:2006at,
Contino:2006qr,
Falkowski:2008fz,
Contino:2010rs,
Contino:2011np} are formulated as simple bottom-up holographic models, in which confinement is modelled by  a hard cut-off in the theory.

\begin{table}[h!]
\caption{Table~I of Ref.~\cite{Elander:2024lir}. 
Field content, organised in terms of  irreducible representations of the symmetries in 
$D=6$ dimensions ($SO(5)$ multiplets),  $D=5$ dimensions ($SO(4)$ multiplets, 
with $\langle {\mathcal X} \rangle \neq 0$), and $D=4$ dimensions ($SO(3)$ multiplets, with $\langle \vec\pi\rangle \neq 0$). In  $D=4$ dimensions, we refer to gauge-invariant combinations, massive representations of the Poincar\'e group.}
\label{Tab:Fields}
\begin{center}
{\footnotesize
\begin{tabular}{|c|c|c||c|c|c||c|c|c|}
\hline\hline
\multicolumn{3}{|c||}{{
$D=6$, $SO(5)$,
}} &
\multicolumn{3}{|c||}{{
$D=5$, $SO(4)$,
}} &
\multicolumn{3}{|c|}{{
$D=4$, $SO(3)$,
}}
\cr
\multicolumn{3}{|c||}{{
{\rm massless irreps.}
}} &
\multicolumn{3}{|c||}{{
{\rm massless irreps.}
}} &
\multicolumn{3}{|c|}{{
{\rm massive irreps.}
}}
\cr
\hline\hline
{\rm Field} & $SO(5)$ & $N_{\rm dof}$  
&{\rm Field} & $SO(4)$ & $N_{\rm dof}$  
&{\rm Field} & $SO(3)$ & $N_{\rm dof}$  \cr
\hline\hline
$\hat{g}_{\hat{M}\hat{N}}$ & $1$ & $9$ &
$g_{MN}$ & $1$ & $5$ &
$g_{\mu\nu}$ & $1$ & $5$ \cr
 & &  &
 & &  &
$g_{\mu5}$ & $1$ & $-$ \cr
 & &  &
 & &  &
$g_{55}$ & $1$ & $-$ \cr
 & &  &
$\chi_M$ & $1$ & $3$ &
$\chi_{\mu}$ & $1$ & $3$ \cr
 & &  &
 & &  &
$\chi_{5}$ & $1$ & $-$ \cr
 & & &
$\chi$ & $1$ & $1$ &
$\chi$ & $1$ & $1$ \cr
\hline
${\mathcal X}_{\alpha}$ & $5$ & $5$ &
$\phi$ & $1$ & $1$ &
$\phi$ & $1$ & $1$ \cr
&&&
$\pi^{\hat{A}}$ & $4$ & $4$ &
$\pi^{\hat{\cal A}}$ & $3$ & $3$ \cr
&&&
&&&
$\pi^{4}$ & $1$ & $1$ \cr
\hline
${\cal A}_{\hat{M}\,\alpha}{}^{\beta}$ & $10$ & $40$ &
${\cal A}_{M}^{\,\,\,\hat{A}}$  & $4$ & $12$ &
${\cal A}_{\mu}^{\,\,\,\hat{\cal A}}$  & $3$ & $9$ \cr
&&&
&&&
${\cal A}_{\mu}^{\,\,\,4}$  & $1$ & $3$ \cr
&&&
&&&
${\cal A}_{5}^{\,\,\,\hat{\cal A}}$  & $3$ & $-$ \cr
&&&
&&&
${\cal A}_{5}^{\,\,\,4}$  & $1$ & $-$ \cr
&&&
${\cal A}_{6}^{\,\,\,\hat{A}}$  & $4$ & $4$ &
${\cal A}_{6}^{\,\,\,\hat{\cal A}}$  & $3$ & $3$ \cr
&&&
&&&
${\cal A}_{6}^{\,\,\,4}$  & $1$ & $1$ \cr
&&&
${\cal A}_{M}^{\,\,\,\bar{A}}$  & $6$ & $18$ &
${\cal A}_{\mu}^{\,\,\,\tilde{\cal A}}$  & $3$ & $9$ \cr
&&&
&&&
${\cal A}_{\mu}^{\,\,\,\bar{\cal A}}$  & $3$ & $9$ \cr
&&&
&&&
${\cal A}_{5}^{\,\,\,\tilde{\cal A}}$  & $3$ & $-$ \cr
&&&
&&&
${\cal A}_{5}^{\,\,\,\bar{\cal A}}$  & $3$ & $-$ \cr
&&&
${\cal A}_{6}^{\,\,\,\bar{A}}$  & $6$ & $6$ &
${\cal A}_{6}^{\,\,\,\tilde{\cal A}}$  & $3$ & $3$ \cr
&&&
&&&
${\cal A}_{6}^{\,\,\,\bar{\cal A}}$  & $3$ & $3$ \cr
\hline
$P_5{}_{\alpha}$&5&5&
$P_5{}_{\hat{A}}$&4&4&
$P_5{}_{\hat{\cal A}}$&3&3\cr
&&&
&&&
$P_5{}_{4}$&1&1\cr
&&&
$P_5{}_{5}$&1&1&
$P_5{}_{5}$&1&1\cr
\hline\hline
\end{tabular}
}
\end{center}
\end{table}

In this proceedings contribution, we  summarize highlights from an ambitious research  programme~\cite{Elander:2021kxk,Elander:2022ebt,Elander:2023aow,Elander:2024lir} (see also Refs.~\cite{Elander:2020ial,Elander:2020fmv,Elander:2021wkc}), which ultimately aims at building a complete holographic CHM model, in which confinement is captured dynamically in the gravity theory. We show how  to combine the spontaneous breaking of an approximate $SO(5)$ symmetry, arising in the background geometry, with weak interactions, localised at the boundary, to induce vacuum misalignment.
We present a simplified, bottom-up holographic model, describing a four-dimensional gauge theory in which a gauged $SO(4)$ subgroup of the $SO(5)$ approximate global symmetry is Higgsed  to its $SO(3)$ subgroup, due to
 misalignment with the vacuum structure of the underlying strongly coupled dynamics. We provide examples of the resulting spectrum, computed using the gauge-invariant formalism developed in
Refs.~\cite{Bianchi:2003ug,Berg:2005pd,Berg:2006xy,Elander:2009bm,Elander:2010wd}---see
also Refs.~\cite{Elander:2009pk,Elander:2012yh,Elander:2014ola,
Elander:2017cle,Elander:2017hyr,Elander:2020csd,Elander:2020fmv,Roughley:2021suu,Elander:2018aub}.
Our results demonstrate the opening up of a (small) hierarchy in the spectrum.
We dispense with the many, non-trivial, technical details necessary in  the construction, which can be found in the extensive, accompanying  publication in Ref.~\cite{Elander:2024lir}. We comment on the next programmatic model-building steps that would lead to a fully realistic model of holographic  CHM with minimal $SO(5)/SO(4)$ coset.

\section{Gravity model and background}

\begin{table}[t]
\small
\caption{Table~II from Ref.~\cite{Elander:2024lir}. Summary table associating the fields in five-dimensional language  to their fluctuations
in the four-dimensional, ADM formalism. 
}
\label{Tab:Fluctuations}
\begin{center}
{\footnotesize
\begin{tabular}{||c|c|c||c|c|c||}
\hline\hline
{\rm Field}  &\multicolumn{2}{|c||}{  {\rm Fluctuation}} 
&
{\rm Field}  &\multicolumn{2}{|c||}{  {\rm Fluctuation}} 
\cr
\hline\hline
$g_{MN}$ &\multicolumn{2}{|c||}{ $\mathfrak{e}_{\mu\nu}$} 
& 
$\left({\cal B}_M{}^{\hat{\mathcal A}},{\cal B}_M{}^{\tilde{\mathcal A}}\right)$ &
\multicolumn{2}{|c||}{$\left(\mathfrak{v}_{\mu}{}^{\hat{\mathcal A}},\mathfrak{v}_{\mu}{}^{\tilde{\mathcal A}}\right)$} \cr
$\chi_M{}$ & \multicolumn{2}{|c||}{ $\mathfrak{v}_{\mu}$} 
&
${\cal A}_M{}^4$ & 
\multicolumn{2}{|c||}{ $\mathfrak{v}_{\mu}{}^4$} \cr
$(\phi,\,\chi)$ & \multicolumn{2}{|c||}{ $(\mathfrak{a}^{\phi},\mathfrak{a}^{\chi})$}
&
 ${\cal A}_M{}^{\bar{\mathcal A}}$ &
 \multicolumn{2}{|c||}{  $\mathfrak{v}_{\mu}{}^{\bar{\mathcal A}}$} \cr
\hline
$\left.\begin{array}{c}
\mathcal B_6^{\hat{\mathcal A}} \cr
 \mathcal A_6^{4}\end{array}\right\}
$ & 
\multicolumn{2}{|c||}{ 
$
\mathfrak{a}^{\hat{A}}=\left\{\begin{array}{c}
\mathfrak{a}^{\hat{\cal A}}
\cr
\mathfrak{a}^{4}
\end{array}\right.$}
&
$\left.\begin{array}{c}
\pi^{\hat{\mathcal A}} \cr
 \Pi^{4}\end{array}\right\}$
& 
\multicolumn{2}{|c||}{ 
$
\mathfrak{p}^{\hat{A}}=\left\{\begin{array}{c}
\mathfrak{p}^{\hat{\cal A}}
\cr
\mathfrak{p}^{4}
\end{array}\right.$}
\cr
$\left.\begin{array}{c}
\mathcal B_6^{\tilde{\mathcal A}} \cr
 \mathcal A_6^{\bar{\mathcal A}}\end{array}\right\}
$ & 
\multicolumn{2}{|c||}{ 
$
\mathfrak{a}^{\bar{A}}=\left\{\begin{array}{c}
\mathfrak{a}^{\tilde{\cal A}}
\cr
\mathfrak{a}^{\bar{\cal A}}
\end{array}\right.$}
&
$$
& 
\multicolumn{2}{|c||}{ 
$
$}
\cr
\hline

\hline\hline
\end{tabular}
}
\end{center}
\end{table}

The bottom-up holographic model described in Ref.~\cite{Elander:2024lir} consists of  gravity in six dimensions coupled to a bulk scalar field, \(\mathcal{X}\), transforming in the (real) vector  representation, ${\bf 5}$, of a gauged \(SO(5)\) symmetry, with gauge field ${\cal A}_{\hat{M}}$,  summarised in Table~\ref{Tab:Fields}. One of the non-compact spacetime dimensions,  \(\rho\), serves as the holographic direction. We focus attention on background solutions with  asymptotically AdS\(_6\) geometry for large values of \(\rho\), corresponding to the ultraviolet (UV) regime of the putative dual field theory. Another spatial dimension is compactified on a circle that smoothly shrinks to zero size at a finite value of the radial direction,  \(\rho = \rho_o\), marking the infrared (IR) regime. The termination of the space at this point introduces a mass gap in the dual field theory, mimicking the effects of confinement in the four-dimensional field theory~\cite{Witten:1998zw}.

In the background solutions, \(\mathcal{X}\) develops a non-trivial profile, spontaneously breaking   the \(SO(5)\) gauge symmetry  to its \(SO(4)\) subgroup.  Furthermore, a boundary-localised (spurion) field, $P_5$, itself transforming as a ${\bf 5}$, acquires a non-trivial vacuum expectation value (VEV), misaligned with $\langle \mathcal{X} \rangle$, so that the symmetry breaks to $SO(3)$. We report here only the information needed to  keep the presentation self-contained and clarify the notation, referring for details to Refs.~\cite{Elander:2022ebt,Elander:2023aow}. The bulk action is
\begin{align}
	\mathcal S_6^{(bulk)} &= \int \frac{\dd^6 x }{2\pi}\sqrt{-\hat g_6} \, \bigg\{ \frac{\mathcal R_6}{4} - \frac{1}{2} \hat g^{\hat M \hat N}\left( D_{\hat M} \mathcal X \right)^T D_{\hat N} \mathcal X - \mathcal V_6
	 - \frac{1}{2} \Tr \left[ \hat g^{\hat M \hat P} \hat g^{\hat N \hat Q} \mathcal F_{\hat M \hat N} \mathcal F_{\hat P \hat Q} \right] \bigg\} \,.
\end{align}
Here, \(\hat{M} = 0, 1, 2, 3, 5, 6\) denote the six-dimensional spacetime indices. The metric in six dimensions, \(\hat{g}_{\hat{M} \hat{N}}\), has a determinant \(\hat{g}_6\) and signature mostly `$+$'. The Ricci scalar for the six-dimensional spacetime is denoted by \(\mathcal{R}_6\). The covariant derivatives are denoted as $D_{\hat N}$, and $ F_{\hat M \hat N}$ is the $SO(5)$ field strength. The scalar potential reads:
\beq
\label{eq:VfromW}
	\mathcal V_6  = -5 - \frac{\Delta (5 - \Delta)}{2} \phi^2 - \frac{5 \Delta^2}{16} \phi^4 \,,
\eeq
in terms of $\phi$, which appears in the parametrisation of the scalar field, \(\mathcal{X}\), as
\[
\mathcal{X} \equiv \exp \left[ 2 i \sum_{\hat{A}} \pi^{\hat{A}} t^{\hat{A}} \right] \mathcal{X}_0 \, \phi, \qquad \text{where} \qquad \mathcal{X}_0 \equiv (0, 0, 0, 0, 1)^T,
\]
with \(\hat{A} = 1, \ldots, 4\), indexing the generators of the \(SO(5)/SO(4)\) coset.
The four PNGBs, \(\vec{\pi} = (\pi^1, \pi^2, \pi^3, \pi^4)\), span the \(SO(5)/SO(4)\) coset~\cite{Elander:2024lir}.

We dimensionally reduce the action to  five dimensions, $ \mathcal S_5^{(bulk)}$, and introduce boundaries at finite values of the radial direction, \(\rho = \rho_i\) for \(i = 1, 2\), which act as regulators. Our calculations are carried out within the constrained range \(\rho_1 \leq \rho \leq \rho_2\), while physical results are recovered in the limits \(\rho_1 \to \rho_o\) and \(\rho_2 \to \infty\). 
Boundary spacetime indices are denoted as \(\mu = 0, 1, 2, 3\)---see Table~\ref{Tab:Fields}.
The complete five-dimensional action, \(\mathcal{S}_5\), contains also boundary terms~\cite{Elander:2024lir}:
\beq \label{eq:bn}
    \mathcal S_5 = \mathcal S_5^{(bulk)} + \sum_{i=1,2} \Big( \mathcal S_{{\rm GHY},i} + \mathcal S_{\lambda,i} \Big) 
        + \mathcal S_{P_5,2} + \mathcal S_{\mathcal V_4,2} 
    + \mathcal S_{\mathcal A,2} + \mathcal S_{\chi,2 } + \mathcal S_{\mathcal X,2} 
\,.
\eeq
where the terms in bracket make the variational problem well defined, while
\begin{align}
\label{eq:SP5}
	\mathcal S_{P_5,2} =& \int \dd^4 x \sqrt{-\tilde g} \, \bigg\{ -\frac{1}{2} K_5 \, \tilde g^{\mu\nu} \left( D_\mu P_5 \right) D_\nu P_5 - \lambda_5 \left( P_5^T P_5 - v_5^2 \right)^2 \bigg\} \bigg|_{\rho = \rho_2} \,,\nonumber \\
	\mathcal S_{\mathcal V_4,2} =& - \int \dd^4 x \sqrt{-\tilde g} \, \mathcal V_4(\mathcal X, \chi, P_5) \bigg|_{\rho = \rho_2} \,,\nonumber\\
	\mathcal S_{\mathcal A,2} \big|_{P_5 = \overline{P_5}} &=
	 \int \dd^4 x \sqrt{-\tilde g} \, \bigg\{ -\frac{1}{4} \hat D_2 \,
\tilde{g}^{\mu\rho}\tilde{g}^{\nu\sigma}{\cal F}^{\hat A}_{\mu\nu}{\cal F}^{\hat A}_{\rho\sigma} 
-\frac{1}{4} \bar D_2 \,
\tilde{g}^{\mu\rho}\tilde{g}^{\nu\sigma}{\cal F}^{\bar A}_{\mu\nu}{\cal F}^{\bar A}_{\rho\sigma} \bigg\} \bigg|_{\rho = \rho_2} \,.\nonumber\\
\mathcal S_{\mathcal X,2} & =
	 \int \dd^4 x \sqrt{-\tilde g} \, \bigg\{ - \frac{1}{2} K_{\mathcal X,2} \, \tilde g^{\mu\nu} (D_\mu \mathcal X)^T D_\nu \mathcal X \bigg\} 
	 \bigg|_{\rho = \rho_2} \,. \nonumber
\end{align}
We set $P_5 = \overline{P_5}$ for simplicity~\cite{Elander:2024lir}. The parameters
$K_5$, $\lambda_5$, $\hat D_2$, $\tilde D_2$, $K_{X,2}$
are discussed later.

\subsection{Model parameters and $SO(4)$ gauging }

The boundary terms in the action, Eq.~(\ref{eq:bn}), are used in the regularisation process, 
implemented along the lines of holographic renormalisation~\cite{Bianchi:2001kw, Skenderis:2002wp, Papadimitriou:2004ap}.
Their finite parts are physical parameters in our analysis. 
The spurion field, $P_5$, is introduced so that all symmetry-breaking effects have spontaneous origin in the gravity formulation~\cite{Elander:2024lir}. The boundary term $\bar D_2$ contains the free parameter, $\bar \varepsilon^2$, that controls the strength of the gauging of the $SO(4)$ in the field theory. In the next section we comment on parameters, $m_4^2$ and $v$,  appearing in the boundary potential, $\mathcal V_4(\mathcal X, \chi, P_5)$. The symmetry breaking pattern $SO(5)\rightarrow SO(4)$ is controlled by $k_{\mathcal X}\equiv K_{X,2}e^{\rho_2(8/3-\Delta)}$~\cite{Elander:2024lir}.

The presence of boundary localised terms breaks the $SO(5)$ symmetry to a gauged $SO(4)$ subgroup, 
which may or may not be aligned to the unbroken $SO(4)$ subgroup, depending on the value of the vacuum misalignment angle, $v$. In the background, this parameterises the non-zero value of $\pi^4 = v$, and leads to the spontaneous breaking of the gauged $SO(4)$ to $SO(3)$. We introduce indices adapted to $SO(3)$, specifically $\hat{\mathcal{A}} = 1,\, 2,\, 3$, $\tilde{\mathcal{A}} = 5,\, 6,\, 7$, and $\bar{\mathcal{A}} = 8,\, 9,\, 10$. These are chosen so that $t^{\bar{\mathcal{A}}}$ represents the unbroken generators of $SO(3)$. The fluctuations of the fourth component of $\pi^{\hat{A}}$ is written as $\pi^4 = v + \Pi^4$. 
In the spin-1 sector, there is  mixing between the two triplets, denoted by the indices $\hat{\mathcal{A}}$ and $\tilde{\mathcal{A}}$. We define the following linear combinations:
\beqs
	\mathcal B_6^{\hat{\mathcal A}} &\equiv& \cos(v) \mathcal A_6^{\hat{\mathcal A}} + \sin(v) \mathcal A_6^{\hat{\mathcal A}+4} \,, \\
	\mathcal B_6^{\tilde{\mathcal A}} &\equiv& - \sin(v) \mathcal A_6^{\tilde{\mathcal A} - 4} + \cos(v) \mathcal A_6^{\tilde{\mathcal A}} \,, \\
	\mathcal B_M{}^{\hat{\mathcal A}} &\equiv& \cos(v) \mathcal A_M{}^{\hat{\mathcal A}} + \sin(v) \mathcal A_M{}^{\hat{\mathcal A}+4} \,, \\
	\mathcal B_M{}^{\tilde{\mathcal A}} &\equiv& - \sin(v) \mathcal A_M{}^{\tilde{\mathcal A}-4} + \cos(v) \mathcal A_M{}^{\tilde{\mathcal A}} \,.
\eeqs
We adopt this basis for the fields (excluding the metric) that  fluctuate around the backgrounds:
\beqs
\label{eq:flucbasis1}
	\Phi^a &=& \{ \phi, \chi \} \,, \\
\label{eq:flucbasis2}
	\Phi^{(0)a} &=& { \{ \mathcal B_6^{\hat{\mathcal A}}, \mathcal A_6^4, \mathcal B_6^{\tilde{\mathcal A}}, \mathcal A_6^{\bar{\mathcal A}} \} } \,, \\
\label{eq:flucbasis3}
	V_M{}^A &=& { \{ \chi_M, \mathcal B_M{}^{\hat{\mathcal A}}, \mathcal A_M{}^4, \mathcal B_M{}^{\tilde{\mathcal A}}, \mathcal A_M{}^{\bar{\mathcal A}} \} } \,, \\
\label{eq:flucbasis4}
	{\cal H}^{(1)}_M{}^A &=& { \left\{ 0, \frac{\sin(v)}{v} \partial_M \pi^{\hat{\mathcal A}} + \frac{g}{2} \mathcal B_M{}^{\hat{\mathcal A}}, \partial_M \Pi^4 + \frac{g}{2} \mathcal A_M{}^4, 0, 0 \right\} } \,.
\eeqs
We use different symbols to distinguish the original fields in the action from the gauge-invariant combinations of fluctuations associated with them---see Table~\ref{Tab:Fluctuations}.

The family of backgrounds of interest is characterised by two parameters: $\Delta$, which is linked to the dimension of the dual field-theory operator responsible for breaking $SO(5)$ to $SO(4)$, and 
$\phi_I=\phi(\rho = \rho_o)$, that  controls the size of symmetry breaking effects. We impose the upper bound $\phi_I \leq \phi_I(c)$, with $\phi_I(c)$ the critical value at which a first-order phase transition occurs---see Ref.~\cite{Elander:2024lir}---and beyond which these solutions would be metastable, and eventually unstable.

The strength of the $SO(4)$ gauge coupling in the dual field theory is approximately $g_4 \equiv \bar \varepsilon g$, where $g$ represents the bulk $SO(5)$ coupling. We restrict attention to small values of the renormalization constant, $\bar \varepsilon$,  to justify the use of perturbation theory.

We can dial the symmetry breaking parameters, $v$ and $m_4^2$, to values that induce the spontaneous breaking of the gauged $SO(4)$ to $SO(3)$, while also producing a separation between the mass scales of parametrically light states and other heavier resonances. The light states, in the four dimensional language, are three massless gauge fields and three massive (but light) vectors, associated with the Higgsing $SO(4)\rightarrow SO(3)$, and one additional scalar singlet.

\section{Numerical results: the mass spectrum}
 
\begin{figure}[t!]
 \includegraphics[width=0.98\textwidth]{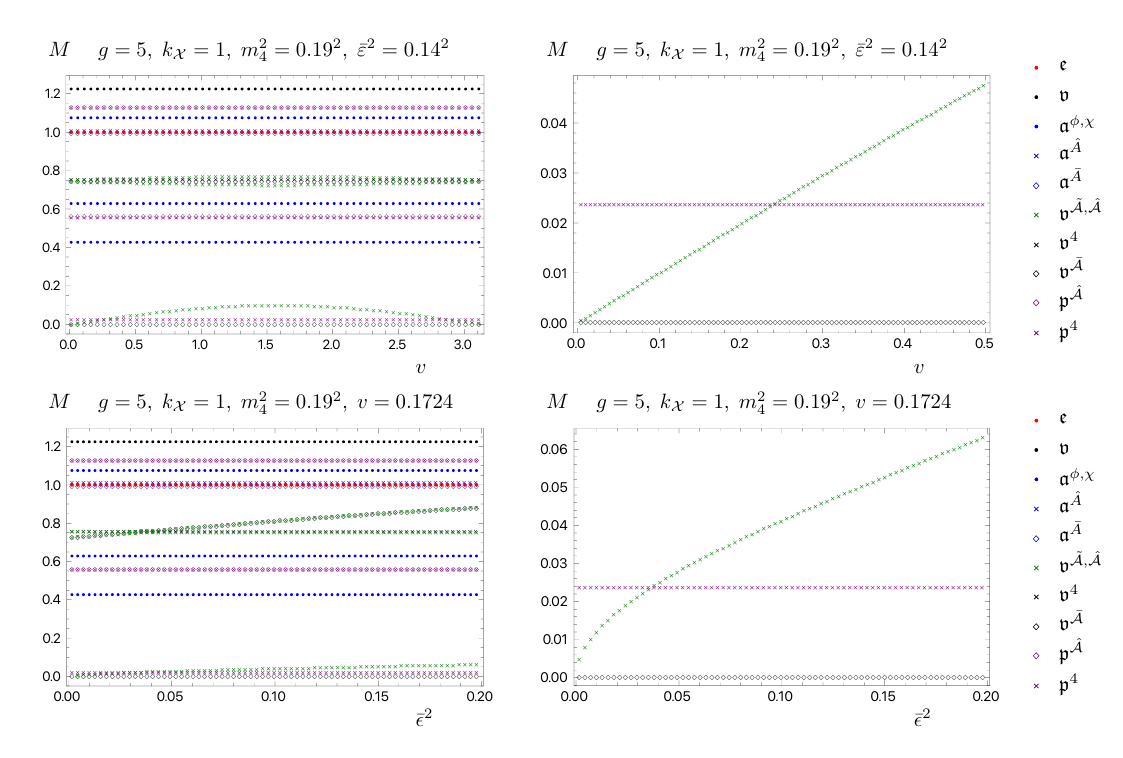}
	\caption{
Figures~3 and~4 of Ref.~\cite{Elander:2024lir}. Mass spectra, $M$,  of  (gauge-invariant) bosonic fluctuations, as functions of the parameter $v$ (top) and  $\bar{\varepsilon}^2$ (bottom). The right panels are details of the left ones.}
	\label{fig:Spectrum1}
\end{figure}

\begin{figure}[t]
	\centering
	\includegraphics[width=\textwidth]{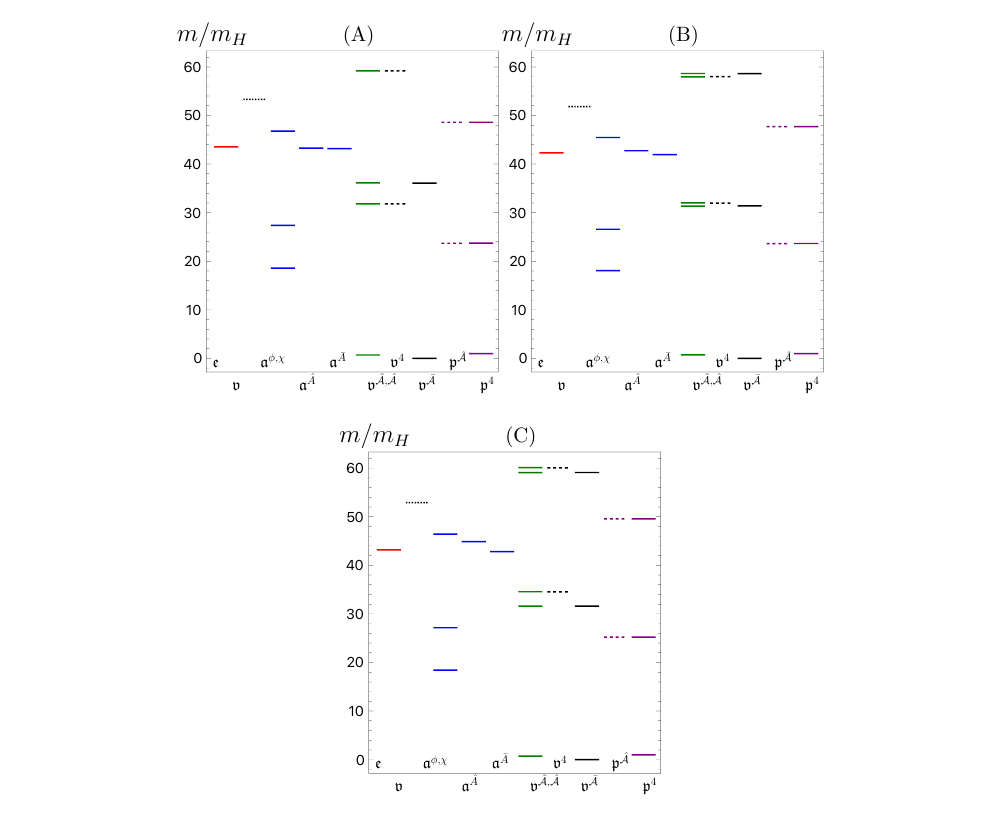}
	\caption{
 Figure~5 of Ref.~\cite{Elander:2024lir}. Illustrative examples of  mass spectra, showing the suppression of the scale of the three lightest states. The masses as normalised to the mass, $m_H$, of the lightest pseudoscalar.}
	\label{fig:Spectrum3}
\end{figure}

We provide examples of the mass spectrum of fluctuations and how they depend on the model parameters. For concreteness, we hold fixed $\Delta = 2$, $\phi_I = \phi_I(c) \approx 0.3882$, $\rho_2 - \rho_0 = 5$, and $\rho_1 - \rho_0 = 10^{-9}$ in this part. Figures~\ref{fig:Spectrum1} and~\ref{fig:Spectrum3} demonstrate how the mass spectrum varies with  $\bar{\varepsilon}$, $g$, $v$, $m_4^2$, and $k_{\mathcal{X}}$. The spectra in Fig.~\ref{fig:Spectrum1}   are normalized to the lightest $SO(3)$-singlet spin-2 fluctuation, $\mathfrak{e}_{\mu\nu}$, while those in 
 Fig.~\ref{fig:Spectrum3}, to the lightest scalar singlet.

Figure~\ref{fig:Spectrum1} displays the spectra for a representative choice of $g$, $k_{\mathcal{X}}$, $\Delta$, $\phi_I$, and $m_4^2$, while varying $v$ and $\bar{\varepsilon}^2$, respectively. The figures reveal several important general characteristics.
Only a few states are light: these include the massless vectors that correspond to zero modes in the unbroken, gauged $SO(3)$ sector, the lightest $\mathfrak{p}^4$ pseudoscalar, and the lightest vectors within the $SO(4)/SO(3)$ coset. All other states have larger masses, demonstrating the opening of a small hierarchy between these two sets of states.
Additionally, the lightest vector states mass increases when either $v$ or $\bar{\varepsilon}^2$ increases, and it approaches zero when either of these parameters is zero. These behaviours are expected on the basis of the fact that these light vectors acquire a mass via the Higgs mechanism.

In Figure~\ref{fig:Spectrum3}, we present three examples, to illustrate what the spectrum of this semi-realistic implementation of CHM looks like. To this purpose,  we denoted by $H$ the lightest mode of the $\mathfrak{p}^4$ fluctuations, and its mass as \( m_H \), and normalised  the other masses against it. We impose the condition \( g_4 = \bar{\varepsilon} g = 0.7 \), to obtain a coupling strength for the $SO(4)$ gauge fields comparable to the \( SU(2)_L \) coupling in the standard model. We then adjust the remaining parameters so that the mass ratio between the lightest fluctuations in the spin-1 sector and the spin-0 sector is approximately \( M_Z/m_H \simeq 0.73 \), reflecting the experimental mass ratio between the \( Z \) and Higgs bosons. 
We show three examples of bosonic spectra that meet qualitative model-building requirements: if we identify the lighter states with experimentally established particles,
with the obvious caveats, we find a rich spectroscopy of new particles appearing with large masses, after a gap in the energy range in which direct and indirect searches for new physics, so far, yield  negative results.


\section{Outlook }

We displayed a semi-realistic 
implementation of vacuum misalignment that meets all the requirements of a CHM, in a context in which calculability extends to include the main properties of heavy,  composite states. The framework we developed, within gauge-gravity dualities, combines the strongly coupled dynamics, captured by the bulk physics of the gravity description, with weak coupling effects, captured by the boundary terms in gravity.
The two next steps of our research programme will take us in opposite, but equally important, directions.
First, to make the phenomenology fully realistic, we would replace the weak gauging of $SO(4)$ with the SM group, $SU(2)\times U(1)$. It would be desirable to also include a treatment of top-quark partial compositeness~\cite{Kaplan:1991dc}, by extending the bulk theory to include fermions in its field content.
Second, as suggested in Ref.~\cite{Elander:2021kxk}, we envision replacing the current, simplified bottom-up gravity action with that of a  known supergravity theory, that can be argued to derive from a 
theory of quantum gravity.
These  challenging, but realistic tasks, are left for the  future.

\begin{acknowledgments}

{\footnotesize

 AF has been supported by the STFC Consolidated Grant ST/V507143/1 and by the EPSRC Standard Research Studentship (DTP) EP/T517987/1.
 MP and AF are supported in parts by the STFC Consolidated Grants No. ST/T000813/1 and ST/X000648/1. MP has also been supported by 
 the European Research Council (ERC) under the European Union’s Horizon 2020 research and innovation program under Grant Agreement No. 813942.

{\bf Open Access Statement}---For the purpose of open access, the authors have applied a Creative Commons 
Attribution (CC BY) licence  to any Author Accepted Manuscript version arising.

{\bf Research Data Access Statement}---The data generated for these proceedings and the extended publication in Ref.~\cite{Elander:2024lir} can be downloaded from  Ref.~\cite{ali_2024}.

} 

\end{acknowledgments}


\bibliographystyle{JHEP}
\bibliography{references.bib}

\end{document}